\documentclass[arguments]{aastex631}

\newcommand{\dsolar}{$\rm d_{\odot}$}

\shorttitle{Origin of the Globular cluste NGC4147}
\shortauthors{Zhang et al.}

\graphicspath{{./}{figures/}}

\begin{document}

\title{Assessing the Association between the Globular Cluster NGC 4147 and the Sagittarius Dwarf Galaxy}

\correspondingauthor{Jundan Nie}
\email{jdnie@nao.cas.cn}

\author[0009-0009-9527-0444]{Ying-Hua Zhang}
\affiliation{Key Laboratory of Space Astronomy and Technology, National Astronomical Observatories, 
Chinese Academy of Sciences, Beijing 100101, P.R China}
\affiliation{School of Astronomy and Space Science, University of Chinese Academy of Sciences, Beijing 
100048, P.R China}

\author[0000-0001-6590-8122]{Jundan Nie}
\affiliation{Key Laboratory of Space Astronomy and Technology, National Astronomical Observatories, 
Chinese Academy of Sciences, Beijing 100101, P.R China}

\author[0000-0003-3347-7596]{Hao Tian}
\affiliation{Key Laboratory of Space Astronomy and Technology, National Astronomical Observatories, 
Chinese Academy of Sciences, Beijing 100101, P.R China}

\author[0000-0002-1802-6917]{Chao Liu}
\affiliation{University of Chinese Academy of Sciences, 100049, China}
\affiliation{Key Laboratory of Space Astronomy and Technology, National Astronomical Observatories, 
Chinese Academy of Sciences, Beijing 100101, P.R China}

\begin{abstract}
The potential association of the globular cluster (GC) NGC 4147 with the Sagittarius (Sgr) dwarf spheroidal
galaxy has been proposed due to their comparable locations and radial velocities. However, 
there are still debates about this connection. In this study, 
we use data from the Dark Energy Spectroscopic Instrument Legacy Imaging Surveys to assess their association. We redetermine the
fundamental parameters of NGC 4147 and find that the cluster is 11.0 Gyr old, has a
metallicity of $Z=0.0006$, and is located 18.5 kpc from the Sun. We utilize the matched filter 
algorithm to identify extratidal structures in the surrounding sky of NGC 4147. The 
multiarmed tidal structures we find align more closely with the result of internal two-body
relaxation processes within the cluster itself. 
The orientations of the dispersed tidal structures, the orbital direction of the
cluster, and the mean orbital direction of Sgr do not show any apparent connection to each
other. It seems to challenge the hypothesis of a common origin between the cluster and Sgr.
To further investigate the association, we study the kinematics of NGC 
4147 with the newly determined fundamental parameters. 
We find that the orbit, orbital energy, and angular momentum of NGC 4147 are not compatible with
those of Sgr or its streams. This suggests that the cluster is not dynamically associated with
Sgr. The morphology and dynamics of NGC 4147 are more consistent with it being a GC that formed with other origin rather than being accreted from the Sgr dwarf galaxy.
\end{abstract}


\section{Introduction} \label{intro}
How do our galaxies form and evolve? This is a complex and active topic of research in
astrophysics. Though some reasonable pictures of these processes have been depicted, there are
still many details to be discovered. As we know, galaxies are made up of many different
components, including stars, gas, dark matter, and stellar clusters. Understanding the composition
and interaction of these components is essential to unravel the diverse properties and behaviors
of galaxies. However, tracing them over time remains a challenging task.  

One of the important tools for probing the host galaxy and tracking galactic history is globular 
cluster (GC) systems. That is because they have survived the changes of their host galaxies over
billions of years and are like witnesses to the entire past of the galaxy 
\citep{brodie2006extragalactic,law2010assessing,forbes2018globular,helmi2018merger,yeh2020origin}.

Galactic GCs, as typical, are the oldest, compact, centrally concentrated, and gravitationally 
bound stellar systems in the galaxy \citep{meylan1997internal}. They have traditionally been
thought to be simple star populations formed from a single giant molecular cloud, with roughly 
uniform age and metallicity. However, modern observations show that GCs have a more complex 
population structure, with multiple age and metallicity distributions 
\citep{2004ARA&A..42..385G,gratton2012multiple,bastian2018multiple}. Thus the attempts to constrain 
the origin of clusters appear more complex.

Generally, GCs in galaxies are thought to have either an in-situ origin or to have been formed
outside the galaxy (ex situ) and deposited by past accretion events. The $\Lambda CDM$ model of 
cosmology predicts that structure formation proceeds hierarchically via mergers. Galaxy mergers 
are a process in which two or more galaxies collide and merge together. This process can bring 
together a large amount of gas, stars, and satellites and is thought to contribute significantly 
to the formation of galaxy halos \citep{johnston1996fossil, stewart2008merger, bland2016galaxy, helmi2018merger,kruijssen2019formation}. 
The convincing evidence comes from the strong correlation of galaxy tidal streams, the remnants or 
relics of these events, and a large fraction of the GCs inside the galaxy \citep{ibata1994dwarf,martin2004dwarf,belokurov2006field,grillmair2009four,crnojevic2016extended}. 

Over time, several satellite accretion events have been identified in our Galaxy, for 
instance, Kraken, the Helmi stream, Gaia-Sausage Enceladus (GSE), Sequoia and the still-accreting 
Sagittarius (Sgr) galaxy (e.g., \cite{ibata1994dwarf,helmi1999debris,chiba2000kinematics,fiorentin2005structure,dettbarn2007signatures,helmi2008stellar,belokurov2018co,helmi2018merger,haywood2018disguise,myeong2019evidence,villanova2019detailed,kruijssen2020kraken,oria2022revisiting,dodd2023gaia,ceccarelli2024walk}). 
The most pronounced accretion event was the spectacular merger of Sgr with the Milky Way. This merger contributed to the formation of the Sgr streams and built up a fraction of the Galactic halo (e.g., \cite{ibata1994dwarf,mateo1996discovery,majewski2003two,newberg2003sagittarius,correnti2010northern,antoja2020all,naidu2020evidence}). 
For now we can convince that there exists some additional GCs that now lie scattered throughout 
the halo may have been stripped from Sgr during its long-term interaction with the Milky Way \citep{majewski2003two,law2010assessing,bellazzini2020globular,minniti2021discovery}. 
Regarding these GCs originally belonging to satellite galaxy Sgr that is still being accreted, 
it is possible to trace their origin by comparing their positions, velocities, and orbits. 
Simultaneously, those of stars belonging to the associated tidal stream can be used to track the 
ongoing accretion \citep{bellazzini2020globular,pagnini2023distribution}. 

The space position of satellite galaxies becomes the priority for searching associated GCs, i.e., 
by searching near the projection on the sky of the theoretical orbit of the Sgr, we notice quite a 
few GCs lie near the Sgr or its streams \citep{jordi2010search,carballo2014search,bellazzini2020globular}. 
Among these GCs, NGC 4147 is an interesting candidate to be a potential member of the Sgr 
galaxy. Actually, this has been extensively debated in the literature, yielding diverse 
conclusions. For example, \citet{bellazzini2003tracing} used an infrared color-magnitude diagram
(CMD) from the Two Micron All Sky Survey to search for stars belonging to the tidal stream 
of the Sgr around NGC 4147 and obtained some statistically significant detections that strongly 
support the idea that NGC 4147 is related to the Sgr Stream. Since \cite{law2010assessing} 
developed a numerical model to reveal the full-sky distribution of the leading and trailing 
arms of the Sgr stream, they found that NGC 4147 lies near the edge of the T1 stream in projection 
and is discrepant from its radial velocity by 86 km s$^{-1}$. While the proper motion (PM) in the 
$\delta$ direction is consistent with the T1 stream, the large uncertainties weakened its 
credibility. They therefore do not conclusively confirm or deny NGC 4147 to have membership in the 
Sgr stream and leave it to be only a weak candidate. \cite{jordi2010search} found a 
two-arm morphology of NGC 4147 and suggested a connection with the extratidal halos and the 
stream of Sgr. These authors considered the association of NGC 4147 to Sgr as still possible but 
with a low statistical confidence \citep{villanova2016spectroscopic}. However,
\cite{sohn2018absolute} rejected the opinion and ruled out NGC 4147 based on the proper motion
measurements using the data from Hubble Space Telescope (HST). Recently, 
\cite{massari2019origin} and \cite{limberg2022reconstructing} both associated NGC 4147 with the GSE. To verify whether NGC 4147 was indeed a former member of the Sgr and 
accreted by the Milky Way, we propose to obtain a complete view of the extratidal structures 
surrounding NGC 4147. If it is from Sgr, the stripped member stars should show some morphological 
connection with the orbit of the Sgr. Additionally, kinematic studies are an important diagnostic 
tool, as accreted clusters should exhibit dynamical coherence with their parent galaxy.

In this work, we use deeper photometric sky survey data from the Dark Energy Spectroscopic Instrument 
(DESI) Legacy Surveys \citep{dey2019overview} to search for extratidal features around NGC 4147 
and discuss the correlation with the Sgr dSph. In Section \ref{DataMethod}, we describe the survey 
data and methods. We analyze the extratidal features, calculate the orbit and energy, and show the 
results in Section \ref{result}. Finally, in Section \ref{summary}, we assess the association and 
summarize the conclusion.

\section{Data and Method} \label{DataMethod}
\subsection{DESI Legacy Surveys Data} 
\label{data}
The DESI Legacy Imaging Surveys are a combination of three public projects: the Dark Energy Camera 
Legacy Survey (DECaLS), the Beijing-Arizona Sky Survey, and the Mayall $z$-band Legacy Survey, which are motivated by the need to provide targets and optical imaging for the DESI 
spectroscopic survey. With the Dark Energy Camera on the Blanco 4m telescope at Cerro 
Tololo Inter-American Observatory, the 90Prime Camera on the 2.3m Bok Telescope at the Kitt Peak 
National Observatory, and the Mosaic-3 camera on the Mayall Telescope, the survey observed 
$\sim$14,000 square degrees of the sky with $|{b}|>20^\circ$. It is sensitive in the 
$\emph{g}$-, $\emph{r}$-, and $\emph{z}$-band with 5$\sigma$ depths of $\emph{g}=24.7$, 
$\emph{r}=23.9$ and $\emph{z}=23.0$ mag, respectively \citep{dey2019overview}. 

We utilize the $g$ and $r$-band data from the latest DR10 data release to carry out the work. We 
constrain the working area to a field of $20^\circ \times 20^\circ$ ($170^\circ<$ R.A. $< 
190^\circ$ and $8^\circ<$ decl. $< 28^\circ$) with NGC 4147 at the center. Because only stellar sources 
are considered, we select point sources classified as ``PSF" (point-spread function) with PSF magnitudes in the range of $16<g<24$ and signal-to-noise 
ratio $> 5\sigma$ in the DESI catalog. We use the reddening value $\emph{E}(\emph{B}-\emph{V})$ provided by the 
dust maps of Schlegel \citep{schlegel1998maps} and the extinction coefficients of 3.214, 2.165 for 
$\emph{g}$, $\emph{r}$ \citep{dey2019overview} to correct the extinction.

\begin{figure}[ht!]
\centering
\includegraphics[scale=0.28]{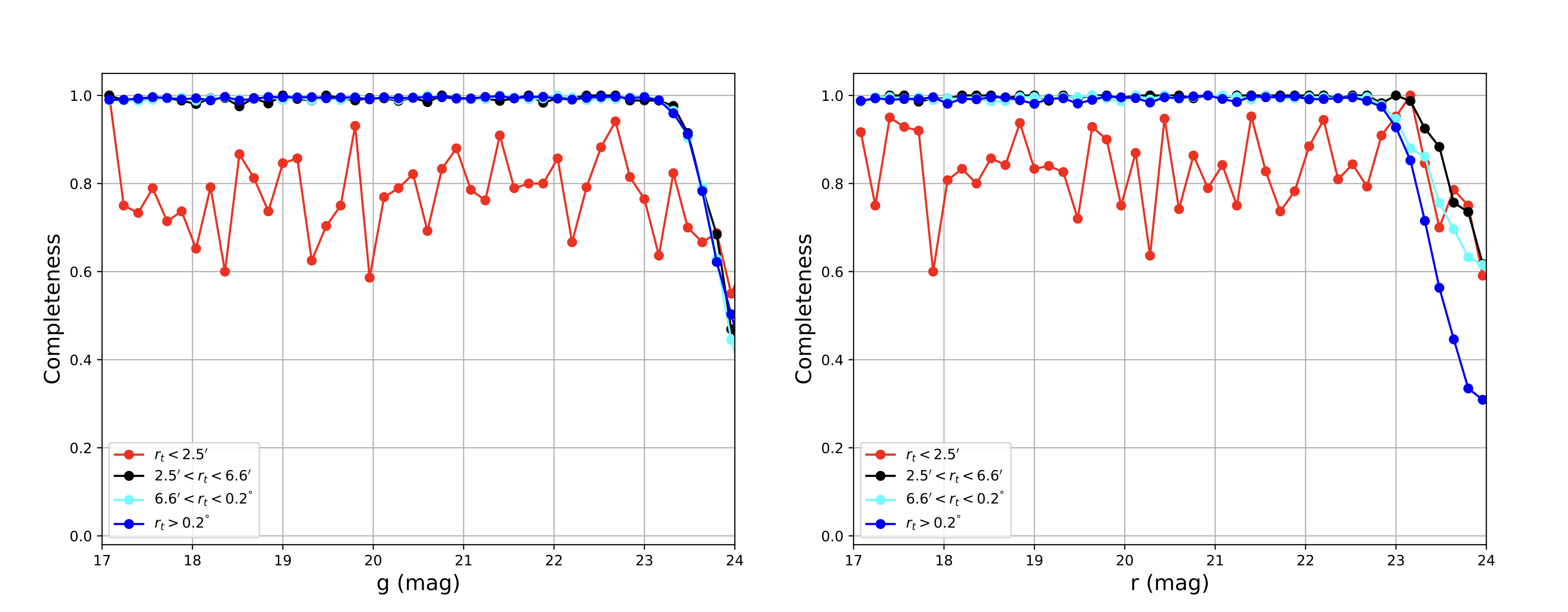}
\centering
\caption{The completeness function for all point sources in our working field. Different colors represent varying distances from the cluster center. The limit magnitudes in $g$ and $r$-band 
are 23.2 and 22.7 mag, respectively.}
\label{figure_depth}
\end{figure}

We perform artificial star tests to evaluate the photometric completeness as a function of magnitude in the $\emph{g}$ and $\emph{r}$-band. Images from the DESI Legacy Imaging Survey centered on NGC 4147 are used for this purpose. Source detection is performed independently by using SExtractor \citep{bertin1996sextractor}. The SExtractor configuration file is debugged to ensure that its source detection and photometry are as consistent as possible with the DESI imaging catalog. We add 3000 artificial stars with random positions and magnitudes from 15 to 24 to these images. All sources are detected by SExtractor and cross-matched with the input artificial stars. This process is repeated 10 times to calculate the average completeness. To reflect the crowding effect, we perform the completeness measurements in four cluster regions: the very dense central region (i.e., $r_t< 2.5^\prime$; see Section \ref{parameters}), a region from cluster center to the tidal radius ($2.5^\prime<r_t< 6.6^\prime$), an intermediate region ($6.6^\prime<r_t< 0.2^\circ$), and a field-star region ($r_t> 0.2^\circ$). The completeness as a function of magnitude and distance from the cluster center is presented in Figure \ref{figure_depth}, with different colors representing different regions. Due to crowding, the central region ($r_t< 2.5^\prime$) has been excluded from our data, so the red lines are provided for reference only. Based on the observed downward trend, we select $\emph{g}=23.2$ mag and $\emph{r}=22.7$ mag as the $100\%$ completeness limit to minimize spatial variations due to survey completeness. These limit magnitudes are 2-3 mag deeper than those other studies have adopted, thus we trust it can reveal more details and provide more convincing evidence.

\subsection{Determination of the Fundamental Parameters of NGC 4147} 
\label{parameters}

According to the GC catalog provided by \citep{harris1996catalog, 
harris2010new}, the cluster NGC 4147 lies at the position R.A. = $12h10m06.2s$, decl. = 
$+18^\circ 32^\prime 31^{\prime\prime}$ $(J2000) $, $l = 253^\circ$ , $b = +77^\circ$ and its
foreground reddening is quite small \citep{stetson2005homogeneous}. The properties of NGC 4147 are
not completely identical from different sources (e.g., \cite{wang2000kinematics,arellano2004ccd,stetson2005homogeneous,villanova2016spectroscopic}), 
while the precise of the fundamental parameters for a cluster, such as age $\tau$, metallicity $Z$, 
and distance modulus $\emph{dm}$ significantly influence the next steps of our work, we 
redetermine the parameters by our own. 

A $\chi^2$ minimization technique is adopted to derive fundamental parameters from the best-fitting isochrone.
The observational data are obtained from DECaLS DR10, cut at $100\%$ completeness with $\emph{g}=23.2$ mag and $\emph{r}=22.7$ mag. This depth is 2-3 mag deeper than the main-sequence turnoff of NGC 4147, allowing for a meaningful 
comparison between the observed CMD in ($\emph{g}-\emph{r}$) versus 
$\emph{r}$ and the theoretical isochrones derived from the PARSEC v1.2+COLIBRI S35 models
\citep{bressan2012parsec,chen2014improving,chen2015parsec,marigo2017new,pastorelli2019constraining}. 
The Salpeter \citep{salpeter1955luminosity} power law is adopted as the initial mass function. 
We calculate the average $E(B–V)$ of the stars around the cluster, and a constant value 
of 0.02 is used.

\begin{figure*}[ht!]
\centering
\includegraphics[scale=0.35]{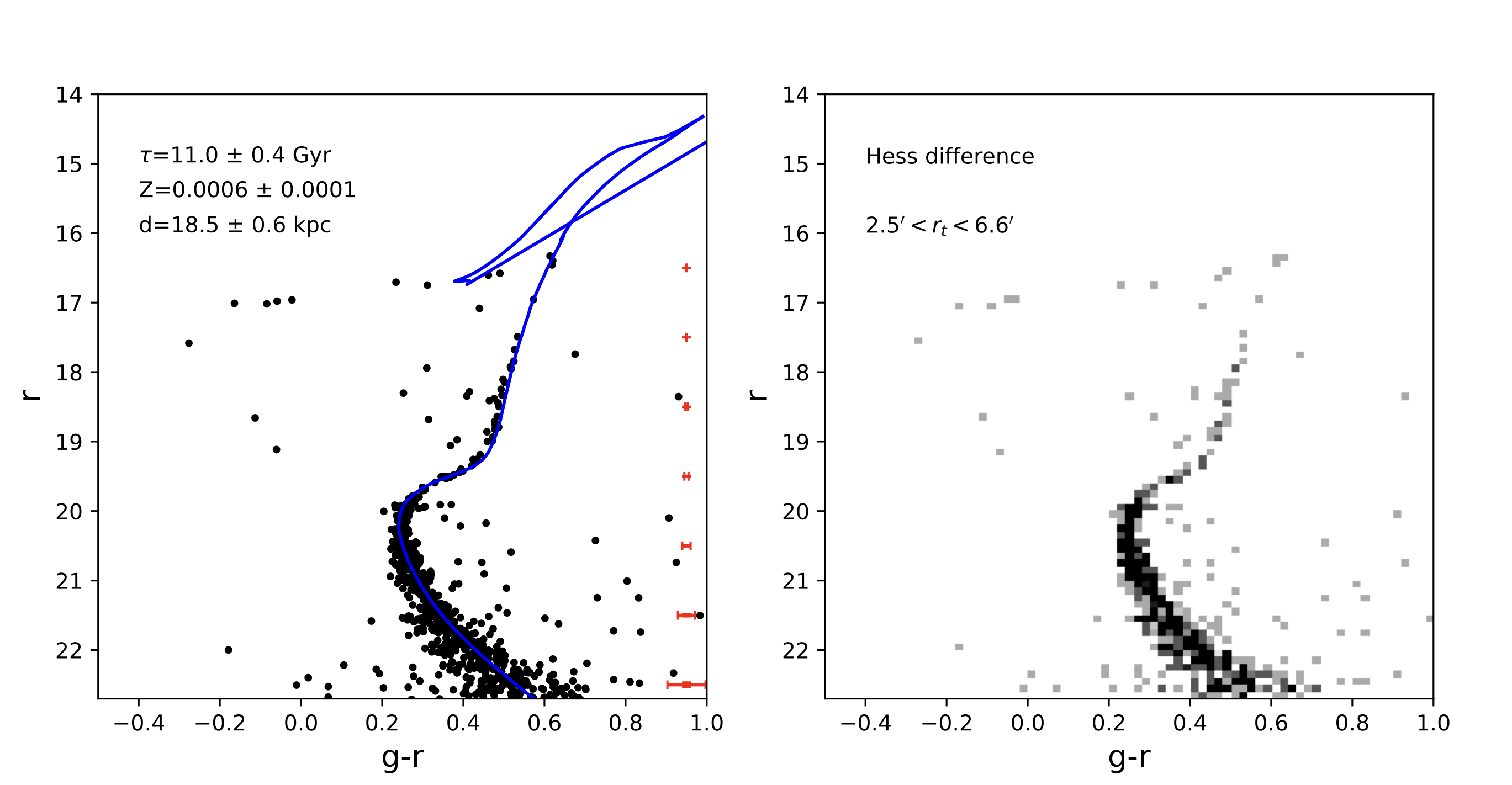}
\caption{\textit{Left panel:} the CMD fitting of NGC 4147. Black dots represent stars 
inside the tidal radius of NGC 4147, excluding the innermost center, i.e., $2.5^\prime < r_t <6.6^\prime$. Typical observational errors are shown from 17 to 22.7 mag with red bars.
The result of the best CMD fitting is $\tau=11.0$ Gyr, Z = 0.0006 and $\emph{\dsolar} =18.5$ kpc (blue 
line).  
\textit{Right panel}: the background-subtracted Hess diagram for NGC 4147,  considered as the final 
CMD template.}
\label{figure_isochrone}
\end{figure*}

The left panel of Figure \ref{figure_isochrone} shows the CMD of stars located within the 
NGC 4147's tidal radius. The tidal radius of NGC 4147 is adopted as $r_{t}=6.6^\prime$ 
\citep{carballo2014search}. To remove stars suffering from crowding, objects inside the very 
center (i.e., $r_t< 2.5^\prime$) are excluded. The crowding size is determined by checking the number density
distribution along luminosity. We perform a Monte Carlo simulation to estimate the best-fitting isochrone using the following approach: First, we generate a Gaussian distribution for each star based on its magnitude and error data ($\mu$,$\sigma$) from DESI DR10. We then draw random magnitude values from this distribution to create a synthetic CMD. By varying the age $\tau$, metallicity Z, and distance modulus $\emph{dm}$ with a moderate range, we calculate the corresponding $\chi^2$ for each ($\tau$,Z,$\emph{dm}$) set and select the isochrone with the minimal $\chi^2$. This process is repeated 1000 times, resulting in 1000 synthetic CMDs and their corresponding best-fit isochrones and parameters. Finally we get 1000 best-fit ($\tau$,Z,$\emph{dm}$) sets, which are represented as Gaussian distributions, with the mean and standard deviation used to determine the parameter values and their associated errors. As shown in 
the left panel of Figure \ref{figure_isochrone}, this best-fitting isochrone has parameters with 
$\tau=11.0\pm0.4$ Gyr, Z = $0.0006\pm0.0001$ (equivalent to [Fe/H] 
= -1.8$\pm0.1$; we adopt an [$\alpha$/Fe] value of 0.38 based on the work of \cite{villanova2016spectroscopic}) and $\emph{\dsolar} =18.5\pm0.6$ kpc ($\emph{dm}$ = 16.33 mag), which is represented by the blue solid line in 
the figure.

We cross-verify our best-fitting isochrone results with prior research:
\begin{enumerate}
\item[.] \cite{stetson2005homogeneous} used ground-based CCD images and HST exposures to create the CMD of NGC 4147. They determined a distance modulus $(m - M)_{V} = 16.40$ 
(equivalent to $\emph{\dsolar}$ = 19.05 kpc) and [Fe/H] = -1.8.
\item[.] \cite{villanova2016spectroscopic} conducted an extensive spectroscopic study of NGC 4147. 
They analyzed the spectra of 18 red giant branch stars and determined a metallicity of [Fe/H] = 
-1.84 $\pm$ 0.02, consistent with typical values for halo GCs in this metallicity 
range.
\item[.] Through Fourier decomposition of RRc stars, \cite{arellano2018variable} calculated 
average metallicity and distance values for the parent cluster NGC 4147 as [Fe/H] = $-1.72 \pm 
0.15$ and $\emph{\dsolar}$ = 19.05 $\pm$ 0.46 kpc.
\item[.] \cite{kruijssen2019formation} determined the mean age and metallicity of NGC 4147 across three samples, reporting a mean age of $12.13\pm 0.46$ Gyr and a mean metallicity of [Fe/H] = 
$-1.66 \pm 0.12$. However, this result is subject to several systematic effects due to the use of different theoretical models. \cite{massari2024euclid}  demonstrated that these discrepancies could easily lead to systematic offsets of more than 1 Gyr.
\end{enumerate}

We get similar results for the fundamental parameters of NGC 4147 within 2$\sigma$ deviation of 
the above parameters. We use these data to calculate the orbit in Section \ref{orbit}.

\subsection{Matched-filter Method} 
\label{method}
\begin{figure}[ht!]
\centering
\includegraphics[scale=0.5]{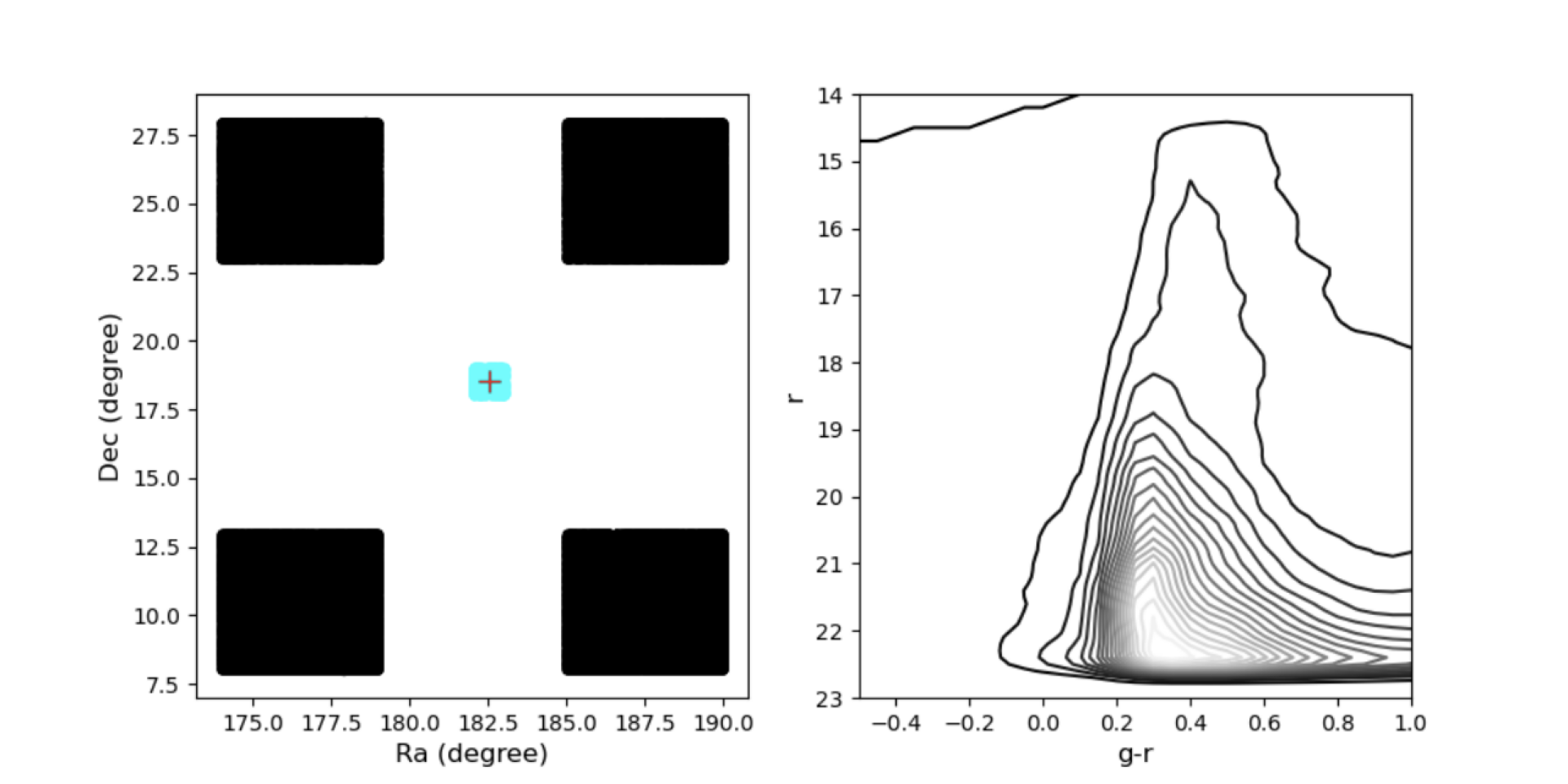}
\caption{The background determination for the MF technique. \textit{Left panel:} The spatial distribution of the four selected background regions. The red cross marks the center of NGC 4147, and the cyan small squares indicate the MF structure regions. \textit{Right panel:} HESS diagram of the background.}
\label{figure_background}
\end{figure}

\begin{figure}[ht!]
\centering
\includegraphics[scale=0.3]{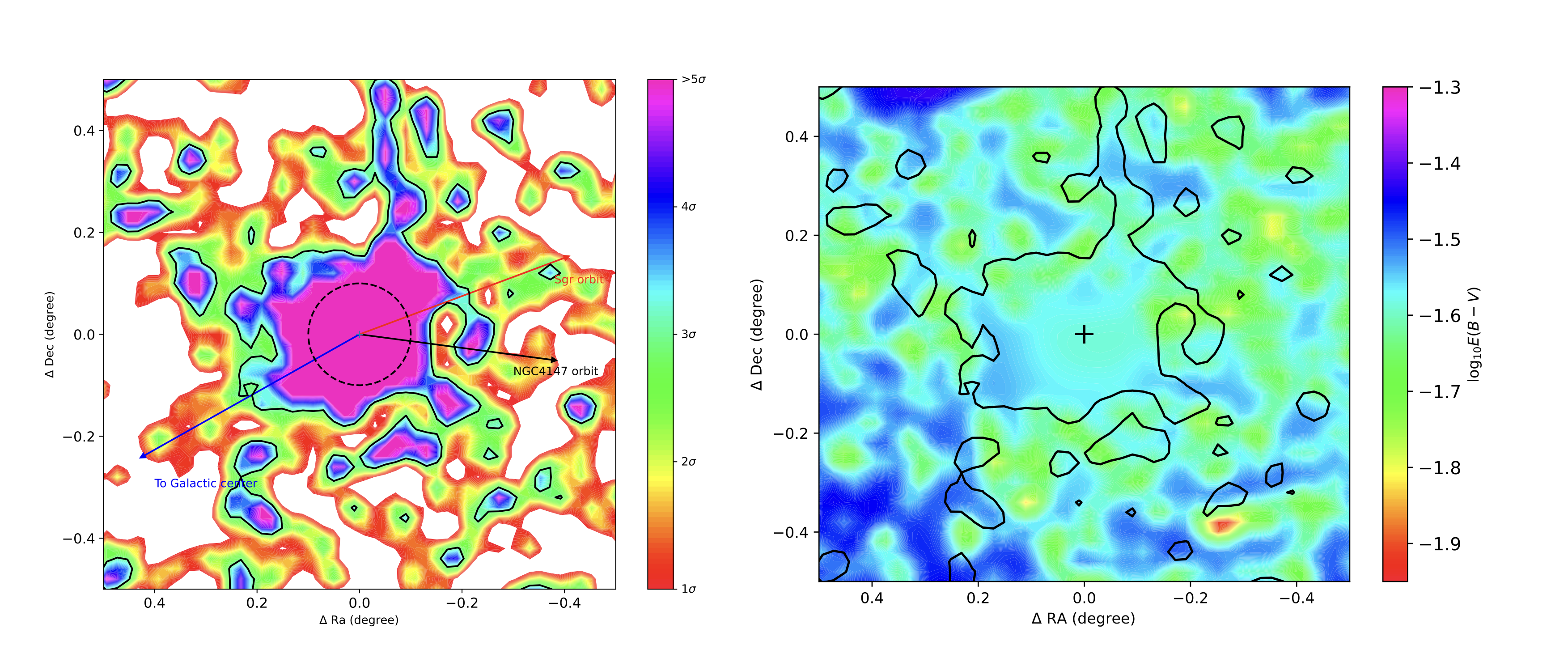}
\caption{\textit{Left panel:} the significance distribution of the density centered by the core of the cluster shows the final 
extratidal structure of NGC 4147. The black dashed circle indicates 
the tidal radius of NGC 4147. Different colors indicate the significance levels with 
a black line for 3.0$\sigma$. The blue arrow indicates 
the direction to the Galactic center, the red arrow for the Sgr orbit, and the black one for the NGC 
4147 orbit. \textit{Right panel:} the extinction map of the same region.}
\label{figure_structure}
\end{figure}

We apply the matched-filter (MF) method to search extratidal features around NGC 4147. The MF is 
developed by Wiener \citep{wiener1949extrapolation}, and applied to GC by \cite{rockosi2002matched}. 
As we consider the $100\%$ completeness in Section \ref{data}, we use partial CMD ($g<=23.2$ mag and $r<=22.7$ mag) as the 
MF template. We constrain the stars within the region of $2.5^\prime < r_t <6.6^\prime$ with $\emph{g}-\emph{r}<1.0$ , 
which led to the field stars being mostly excluded. Then we mimic potential contaminations by
substracting stars in a comparing region, and the Hess difference is considered as the final CMD 
template, as shown in the right panel of Figure \ref{figure_isochrone}. To get the background CMD 
template, we select four sky regions away from the cluster. These regions are rectangular, each with dimensions of $5^\circ$ by $5^\circ$. The number of stars in each background region is normalized according to the area of each region. The background CMD template is binned with the same bin size and color-magnitude bins as the cluster CMD template. The background region and HESS diagram are shown in Figure \ref{figure_background}. The density of the stars 
that passed the MF selection, $\alpha$, is the output of the MF method.  It indicates the final 
extratidal structure of NGC 4147, as plotted in the left panel of Figure \ref{figure_structure}. To avoid any potential impact of extinction on the detected structure, we compare it with an extinction map in the right panel.

\section{Result} \label{result}
\subsection{Extratidal Features around NGC 4147} 
\label{feature}
We obtain the MF output $\alpha$, which is the distribution of the numbers of stars that satisfy the 
CMD template. The background fluctuations are calculated, and the significance of density enhancements is displayed in the left panel of Figure \ref{figure_structure}. This field covers an area of $1^\circ\times 1^\circ$, with black lines indicating 3.0$\sigma$. The influence of extinction has been excluded, as there are no significant regions of high or low extinction correlated with the cluster's structure, as shown in the right panel of Figure \ref{figure_structure}.

We discover a multiarm substructure outside the tidal radius of NGC 4147, with a 
significance higher than 3$\sigma$. We do not detect any evidence of a tidal stream or analog. 
Compared to the findings of \cite{jordi2010search}, which reveal a complex multiple arm 
morphology with a significance above 1$\sigma$, we have two arms in common (loaded at the top and left-hand
direction of the figure). However, the long continuous arm they present in the downward direction seems 
less discernible to us. This could be attributed to different significant levels. If we compare the 
structures at the same significance level, e.g., 1$\sigma$ or 3$\sigma$, we not only confirm their detections but also
uncover even more extensive extensions.  Additionally, the 3$\sigma$ structure we find also overlaps with 
the distribution of extratidal tail members detected by \citet{kundu2022extra}, but ours extends further 
in the downward direction compared to theirs. All the extensions mentioned above arise from the utilization of 
deeper DESI Legacy Surveys data.

We have also compared our findings with a simulation 
predicting the tidal tail of NGC 4147, as 
outlined in the study by \citet{el2022effect}. This study depicts elongated tidal tails 
extending along the two sides of the cluster, a feature not readily apparent in our observations. As 
noted in this study, observing such extended tidal tails in reality may pose a challenge, given that
the simulation assumes a long cluster disruption period of 5 Gyr to generate these features for analysis, which may 
not align with observed phenomena. Due to these variations, we are unsure about the true morphology of NGC 4147 in simulations and 
the extent to which it is observable.

\begin{figure*}
\centering
\includegraphics[scale=0.4]{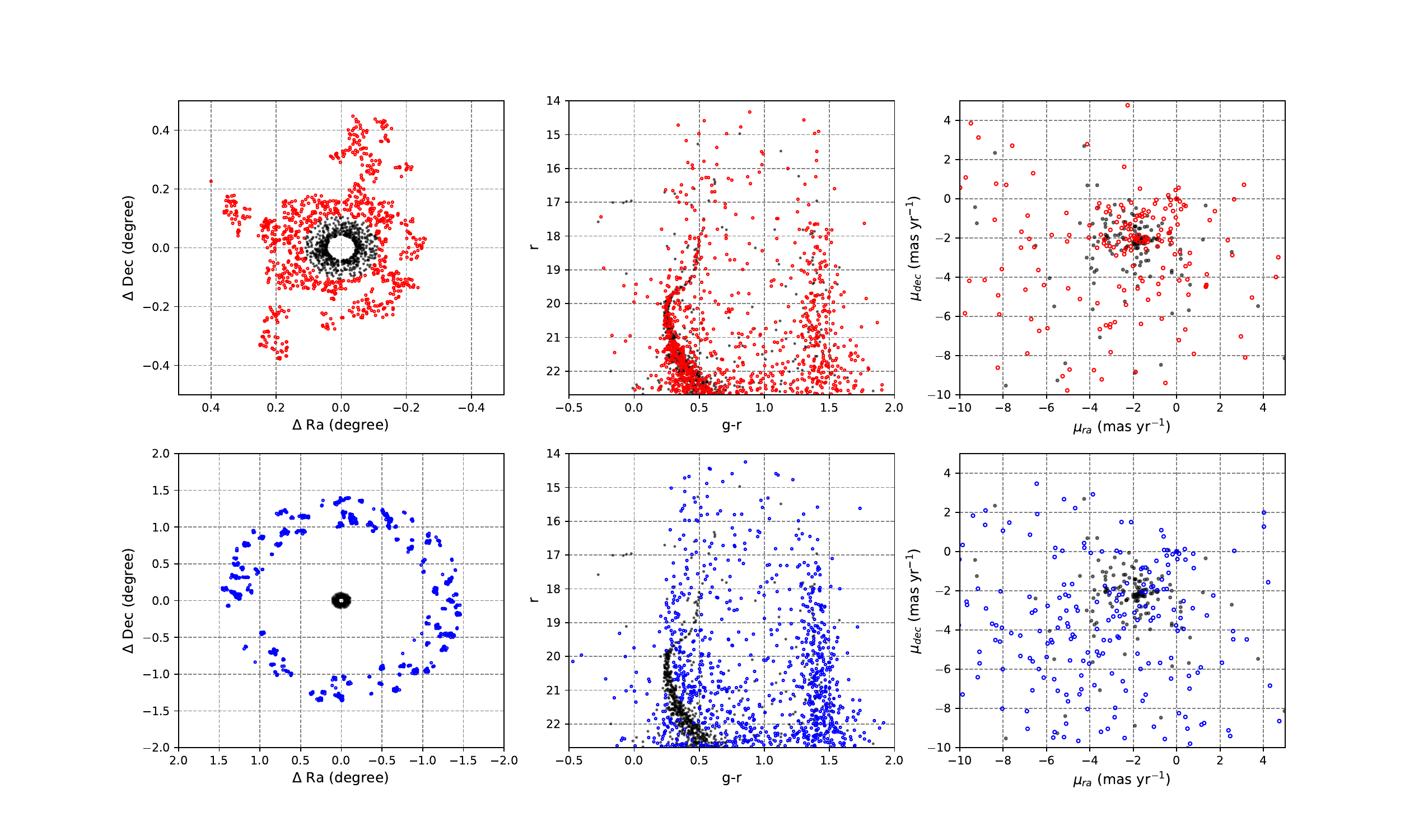}
\centering
\caption{\textit{Left panels:} the comparison of stars' position inside and 
outside the structure of NGC 4147. \textit{Middle panels:} the comparison of the CMD. 
\textit{Right panels:} the comparison of the PM diagram. In all panels, black dots denote stars 
inside the tidal radius ($2.5^\prime < r_t <6.6^\prime$), red circles denote all stars inside the multiarmed structure
(and $r_t >6.6^\prime$), and blue circles denote comparing stars outside the structure (i.e., region without MF signal).}
\label{figure_stars}
\end{figure*}

To verify the reality of our detected feature, we examine stars both inside 
and outside the structure. The upper-left panel of Figure \ref{figure_stars} displays the 
structure stars in regions outside the cluster's tidal radius, represented by red circles. The upper-middle part of Figure \ref{figure_stars} shows the CMD for these structure stars, while the upper-right panel depicts the PM diagram. The lower panels compare this with a nearby field where no 
overdensity (i.e., no MF signal) is detected, represented by blue circles. The number of stars in this comparison region is equal to the number of stars in the structure. The black dots indicate the stars inside the tidal radius (2.5$^\prime<r_t<6.6^\prime$). From the comparison in the middle panels, a clear distinction becomes apparent: the 
majority of structured stars align with the CMD of stars inside the tidal radius, whereas the CMD 
of comparing stars does not overlap with that of stars within the tidal radius. A similar trend can
also be observed in the PM diagram: the red circles and black dots exhibit more overlap and are
distributed more densely (note that not all stars in the middle panels possess PM values due to the
magnitude limit of Gaia PM), while the blue circles distributed more dispersedly on the PM diagram. 
These two comparisons effectively illustrate that the structure stars resemble those of the GC core 
stars, suggesting the credibility of our detected tidal structures.

It should be noted that the structured region includes field stars ($\emph{g}-\emph{r}$ $>$
1.0) since no cut was applied in the top-left panel. This can also account for the PM dispersion in the
top-right panel. Additionally, as depicted in the
middle panel, the majority of stars have magnitude fainter than $G$=19 mag, resulting in decreased PM
accuracy with fainter magnitudes. Therefore, the PM distribution serves solely as a reference,
indicating that the red-circle stars appear more compact and have more overlaps than the blue-circle
stars in the PM diagram.

The identification and analysis of these tails, as well as their shape, extension, orientation, and 
stellar content, provide a wealth of information on the cluster's dynamical interaction and 
evolution. An interesting phenomenon we find in this structure is that it has multiple arms, 
unlike Pal 5 or other GCs, which typically have symmetric s-shaped tidal tails. Instead, there are
multiple arms in the field, and the orientation of the arms is not consistent with 
the direction to the Galactic center (blue arrow in Figure \ref{figure_structure}). 
This suggests that the arms are not primarily created by the tidal forces. Additionally, the 
orientation of the arms, as well as the orbital direction of the GC (black arrow in Figure 
\ref{figure_structure}), does not align with the direction of the Sgr orbit (red arrow in Figure 
\ref{figure_structure}). This observation is different from the findings in \cite{nie2022searching} and
\cite{carballo2017southern}, where Whiting 1, being a member of Sgr, displays a tidal extension direction entirely consistent with 
its own orbit and that of Sgr. The orientation of the tidal extension might 
be related to the orbit phase of the cluster \citep{el2022effect}, however, the inconsistency between the orbital direction 
of the cluster itself and Sgr seems to suggest that NGC 4147 does not share similar kinematics with Sgr.

NGC 4147 is currently located 18.5 kpc from the Sun, near apogalacticon ($R_{apo}$ = 25.5 
kpc), as noted in \cite{dinescu1999space}. The true distance modulus between NGC 4147 and the core 
of the Sgr is $-0.7 \pm 0.15$ \citep{bellazzini2003tracing} or $7.24 \pm 0.5$ kpc. Given the 
relatively weak gravitational potential from both the Galactic center and the Sgr core at these 
distances, it is suggested that the diffuse extratidal features observed in NGC 4147 may be 
primarily influenced by the two-body interactions experienced by the stellar population. The 
two-body relaxation is driven by the gravitational interactions between individual stars in the host 
cluster and surrounding galaxies. This dynamical evolution process can alter the angular momentum 
of stars at large distances, placing them into nearly radial orbits and driving them to 
disruption. Two-body relaxation causes loosely bound stars to gain velocity higher than the 
escape velocity, leading to their evaporation from the cluster. As a result of this interaction, 
stars escape from the center and diffuse around the GC. Owing to this,  
the orientation of the diffuse arms does not align with the orbital direction of the cluster.

Actually, some hint of two-body relaxation is seen from the work by \cite{kundu2022extra}. The authors 
identified 11 extratidal candidates around NGC 4147. However, it was observed that six of these candidates
were located beyond the Jacobi radius of NGC 4147, suggesting that they were completely outside the 
gravitational potential of the cluster. This indicates a possibility that these diffused stars have been 
detached from the cluster.

\begin{figure*}
\centering
\includegraphics[scale=0.28]{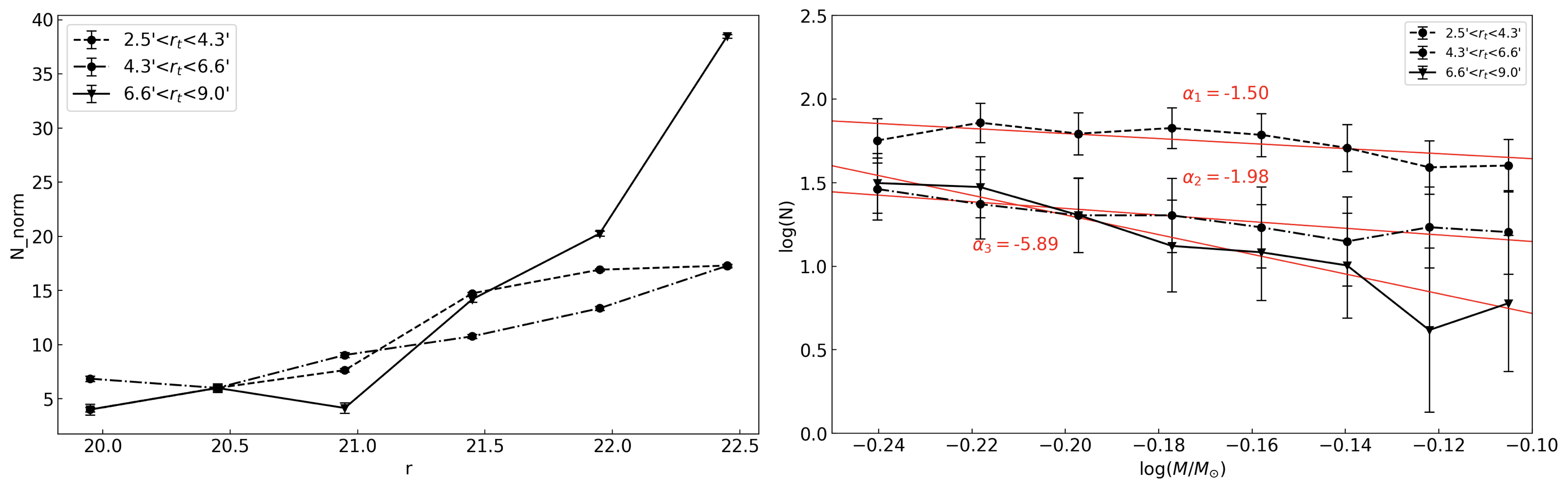}
\centering
\caption{\textit{Left panel:} LFs in three different regions: $2.5^\prime < 
r_t <4.3^\prime$, $4.3^\prime < 
r_t <6.6^\prime$, and $6.6^\prime < r_t < 9.0^\prime$.  The star count is normalized at r = 20.45 mag. \textit{Right panel:} mass functions of NGC 4147. The error 
bars are 1$\sigma$ Poisson error derived from number counts. The solid lines show the best-fit 
power laws. }
\label{figure_relaxation}
\end{figure*}

To find more evidence for the two-body relaxation effect, we explore the luminosity function (LF) and mass function of stars in the NGC 4147 core and its tidal structure. We selected three regions, both inside and outside the tidal radius, to study whether the stars in the structure exhibit significant mass segregation. Within the cluster's tidal radius, we divide the region into two bins. The division point is randomized within the radius range of $2.8^\prime < r_t < 5.5^\prime$, and similar results are found for both bins inside the tidal radius. Ultimately, we choose $r_t = 4.3^\prime$ as a reasonable division point between the two bins ($2.5^\prime < r_t <4.3^\prime$ and $4.3^\prime < r_t <6.6^\prime$). For the outer region, we adopt the range $6.6^\prime < r_t <9.0^\prime$ and select stars located within 3$\sigma$ significance of the MF structure. Accounting for incompleteness and field contamination, we use stars with $r < 22.7$ mag to ensure completeness. Then we select stars within the 2$\sigma$ range of the best-fit isochrone obtained in Section \ref{parameters} on the CMD diagram. These stars are considered to be former members of NGC 4147. Additionally, we choose a region far from NGC 4147 and its tidal structure (i.e., $36^\prime < r_t <48^\prime$) to represent the field stars and select stars within the same range on the CMD diagram. We define the LF in the $r$-band as 
\begin{equation}
    LF_i = N_i - \frac{A_i}{A_{field}}N_{field}
\end{equation}
Here, $N_{i}$ represents the number of member stars per bin in the $r$-band, and $A_i$ is the corresponding area of the cluster regions. The $N_{field}$ refers to the number of field stars. $LF_i$ represents the decontaminated LF. The errors in the counts are assumed to follow a Poisson distribution. Conversion from apparent to absolute magnitudes is obtained using a distance modulus from Section \ref{parameters}, as a distance modulus $dm = 16.33$. 
The left panel of Figure \ref{figure_relaxation} shows the resulted main-sequence LFs for different regions of NGC 4147. The LFs are shifted to r = 20.45 mag to enable a clearer comparison. The LF for the outermost region is much steeper than the inner region, which indicates a significant mass segregation. The MF of NGC 4147 is constructed by converting the LF using the mass–luminosity relation ($M/L$ relation). This $M/L$ relation is derived from a theoretical PARSEC isochrone with fundamental parameters obtained in Section \ref{parameters}. The MFs are presented in the right panel of Figure \ref{figure_relaxation}. The MF slope is almost flat within the tidal radius but exhibits a negative slope outside this radius. This suggests significant mass loss among low-mass stars in the cluster's central region, with lower-mass stars increasingly congregating toward the outer regions. The variation in MF slopes indicates mass segregation, implying a higher concentration of massive stars in the cluster's inner region and a prevalence of low-mass stars in the outer region. This observed mass segregation in the GC is believed to result from its dynamical evolution due to two-body relaxation.

\subsection{The orbit with sgr}
\label{orbit}
As indicated in Figure 4, the orbital direction of the cluster is inconsistent with that of Sgr. Inspired by this, we carefully examine the orbits of the two systems in this section.

With the assistance of the N-body model for the tidal disruption of Sgr (LM10 hereafter; \cite{Law2010THESD}), which provides a reasonably good description of the position and kinematics of the stars lost more recently by Sgr, we are able to further investigate the association between NGC 4147 and Sgr. 
The 6D parameters of Sgr streams are obtained from the simulation of the LM10 model. 

For NGC 4147, the distance is redetermined in Section \ref{parameters} while the PM data are from \cite{vasiliev2021gaia}'s work.
The radial velocity is set to 179.35 $\pm$ 0.31 km s$^{-1}$ \citep{baumgardt2018catalogue}. Together with all 
the 6D information requested, we use the orbit function in \tt{Galpy}\normalfont{} 
\citep{bovy2015galpy} package to trace back the orbit of the cluster, assuming the \tt{MWPotential2014} 
\normalfont{} potential \citep{bovy2015galpy}. The potential includes three parts: the bulge, the 
disk, and the halo. The orbit of NGC 4147 has been computed back in time for approximately 0.8 Gyr.

\begin{figure*}[ht!]
\centering
\includegraphics[scale=0.25]{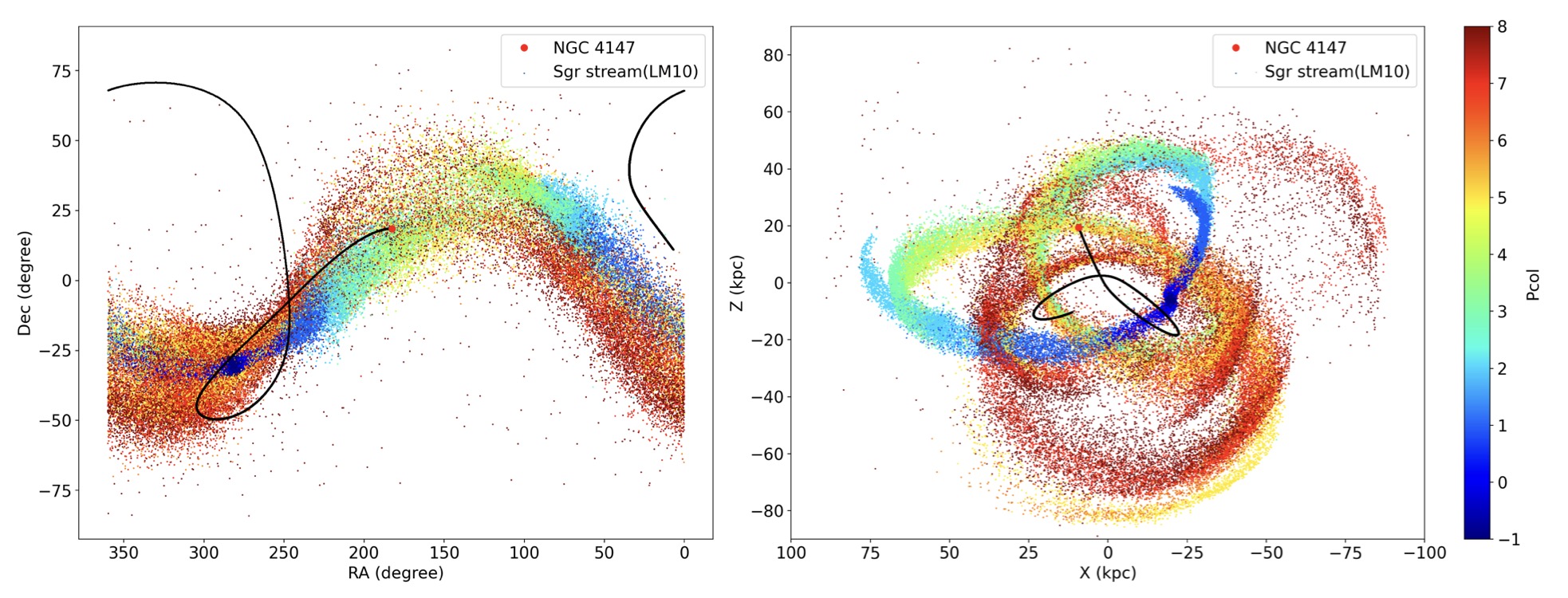}
\caption{The orbits of Sgr stream model particles and NGC 4147. \textit{Left panel:} the colored dots represent 
the Pcol (from the LM10 model), while the 
black line shows the orbit of NGC 4147 in the R.A.-decl. planes. \textit{Right panel:} in the X-
Z planes.}
\label{figure_orbit}
\end{figure*}

Our results are shown in Figure \ref{figure_orbit}. The left panel displays the orbit coordinate positions of NGC 4147 (red dot and black line) and the model particles of the Sgr streams (colored dots for different Pcol) in the R.A.-decl. planes, while the right panel displays them in the X-Z planes. Pcol represents the perigalactic passage when the model particles were stripped from the Sgr. Pcol = -1 indicates that the particles are still gravitationally bound to the main body of the Sgr. Conversely, Pcol = 0, 1,..., 8 represents the current perigalactic passage and one up to eight perigalactic passages ago.
\cite{bellazzini2020globular} recommended NGC 4147 as a candidate due to its similarity in spatial position and radial velocity in the galactocentric reference frame. However, we can intuitively conclude that, despite the projection of their locations, NGC 4147 and the 
Sgr streams have followed very different trajectories since ancient times. 
Therefore, the disagreement between their orbit path supports no dynamical association of NGC 4147 and Sgr.

\subsection{The total orbital energy and angular momentum plane} 
\label{energy}
\begin{figure}[ht!]
\centering
\includegraphics[scale=0.4]{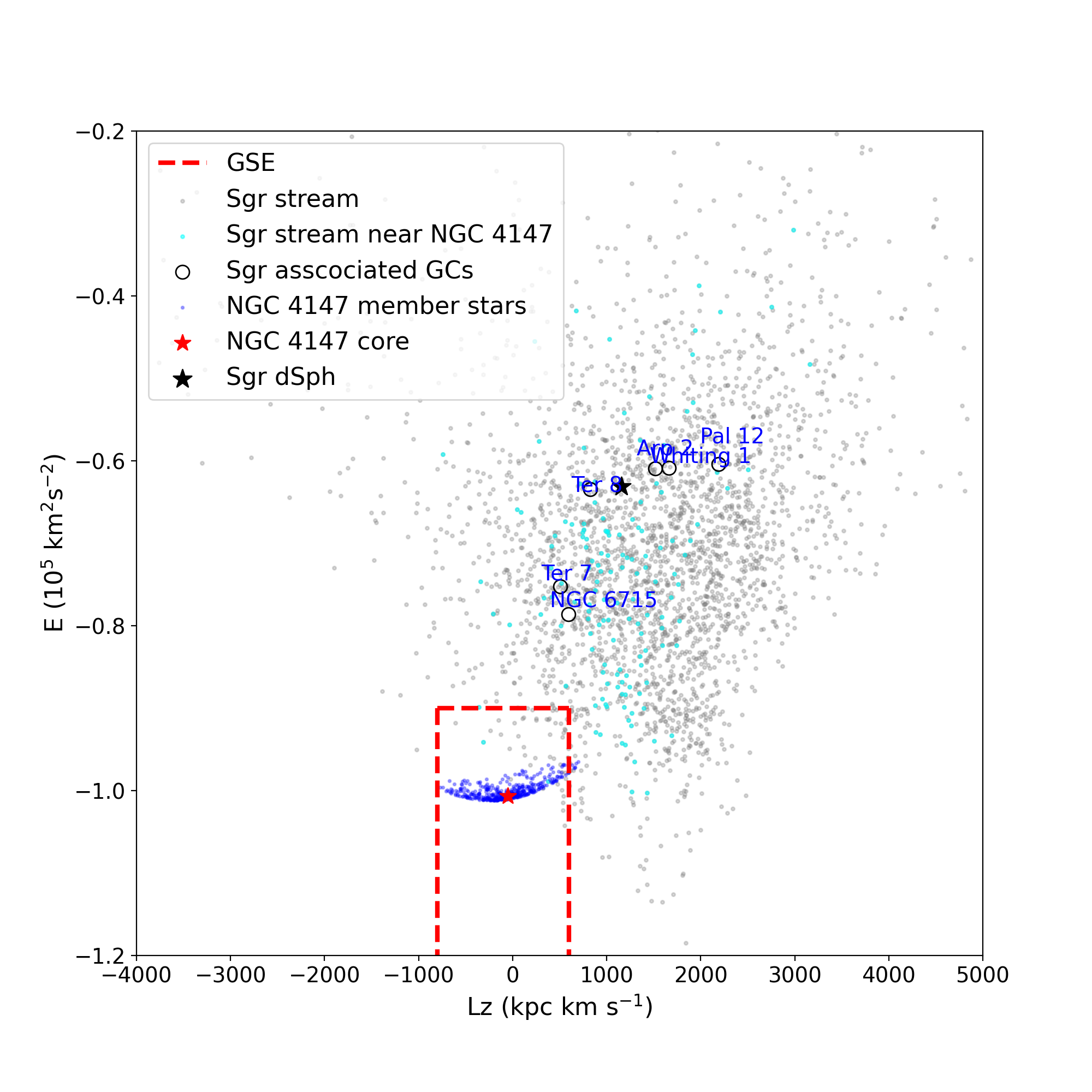}
\caption{Relationships between the angular momentum along the z-direction Lz vs. total orbital energy 
E. The gray dots show the distribution of Sgr stream stars, and the cyan dots are the ones near NGC 4147 in space position. The black star represents the Sgr dSph core and the red one NGC 4147. The black circles are the GCs associated with Sgr. The blue dots stand for the high-probability member stars of NGC 4147. The red dashed box denotes the region of concentrated energy-angular momentum within the GSE (collected from \cite{massari2019origin}).}
\label{figure_energy}
\end{figure}

To further investigate the association between Sgr and NGC 4147, we construct a plot of 
angular momentum along the z-direction (Lz) versus total orbital energy (E). We use observed orbital 
parameters for Sgr streams \citep{yang2019tracing} and for GCs that have been proposed as potential 
accretions from Sgr. We choose six GCs that were suggested as systems formed in the interior of 
the Sgr dSph and latter accreted by the Milky Way: M54 (NGC6715), Arp2, Ter7, Ter8, Pal12 
and Whiting 1 (e.g., \cite{bellazzini2003tracing,carballo2014search,baumgardt2019mean,law2010assessing,arakelyan2020globular,minniti2021discovery,nie2022searching}). 
As shown in Figure \ref{figure_energy}, these associated GCs exhibit a similar distribution in the 
Lz-E plane, with Lz typically falling between -60000 and -80000 km$^2$ s$^{-2}$, similar to the orbital energy 
range of the Sgr dSph itself. On the contrary, NGC 4147 possesses smaller Lz and E, placing it outside the main distribution of Sgr. Despite having a similar age and 
metallicity to Apr 2, NGC4147's Lz-E location implies that it has different dynamics relative to 
Sgr and other Sgr member GCs. Moreover, the Lz-E distribution of NGC4147 also differs from the 
stars in the vicinity of NGC 4147 that are affiliated with Sgr (as shown in cyan in Figure 
\ref{figure_energy}). This supports the conclusion that NGC 4147 lacks kinematic resemblance to 
Sgr. We argue that if NGC 4147 was associated with Sgr or formed from the merger 
event of Sgr, it should exhibit similar kinematic behavior to other accreted GCs. This is 
because they share a same progenitor, which gravitationally bound them together. Therefore, when 
the merger occurred, their collective behavior should have been preserved and shown in their 
kinematics.

Our work arrives at a similar conclusion to \cite{arakelyan2020globular}, who analyzed the
associations between GCs and the Sgr tidal stream by examining spatial positions, radial 
velocities relative to the galactic standard of rest, proper motions, and ratio of age to 
metallicity. In their assessment, NGC 4147 was identified as one of the lowest-ranked candidates 
in terms of association with Sgr.

\cite{massari2019origin} suggested that NGC 4147 might have originated from GSE. Based on 
their Figures 2 and 3, which show the Lz-E plane for a large sample of GCs, we have
marked their Lz-E range of the GSE on our plot (the red box in our Figure \ref{figure_energy}). 
As seen from the figure, NGC4147 falls precisely within this region. Additionally, 
\cite{limberg2022reconstructing} identified NGC 4147 as a 
confident GSE member, with a final membership probability exceeding 70 percent.

However, the association between NGC 4147 and GSE is not certain and requires further discussion. 
This remains the quest to understand whether the GC NGC 4147 actually formed in situ 
or in progenitors of GSE accretion events. The age of NGC4147 we obtained in Section 
\ref{parameters} is relatively young, which may further support the idea that NGC 4147 could have 
had an accreted origin \citep{marin2009acs}. That is, the possibility of it forming ex-situ remains more plausible. 

In conclusion, based on our analysis, we believe that NGC 4147 is not an association of Sgr. It 
could be either an outer halo intruder or an accreted cluster member of the GSE event, although 
the latter is more likely. To verify its origin, it will be crucial to 
obtain chemical components and radial velocity for more member stars, including extratidal stars of NGC 
4147, using spectroscopic surveys such as the China Space Station Telescope and DESI. 
Our future work will therefore focus on investigating the element trend using detailed 
chemical abundance analysis of spectroscopic data. By extending this study to other GCs, we can
develop a deeper understanding of how GCs formed and were accreted into our Galaxy.

\section{Summary} 
\label{summary}
We explore the origin of NGC 4147 from two aspects: morphology and kinematics. First, we 
redetermine the fundamental parameters of NGC 4147 (age, metallicity, and distance) by finding the 
best-fitting isochrone. To detect extratidal features around NGC4147, we use the deep DESI Legacy 
Survey data, taking into account the completeness and depth of the data. We constrain a work area 
to a field of $15^\circ \times 10^\circ$ and apply the MF method to select stellar objects within 
the area. We arrive at the following results:
\begin{enumerate}
\item[1).] A multiple arm substructure outside the tidal radius of NGC 4147 has been detected with a 
significance higher than 3$\sigma$; 
\item[2).] The substructure shows a remarkable diffuse distribution. The orientation of the
structure extension
matches neither the direction to the Galactic center nor the Sgr orbit. Additionally, the orbital direction 
of the cluster does not align to the orbital direction of the Sgr.
These indicate no apparent morphological connection between NGC4147 and the Sgr.
\item[3).] A mass segregation between the core and the outer regions of the cluster has been identified, supporting that the diffuse structure is a result of two-body relaxation. 
\end{enumerate}

Furthermore, kinematics strengthen some of the results above: 
\begin{enumerate}
\item[1).]The calculation of orbit shows the projection of location between NGC 4147 and the Sgr 
streams coincides for a very short period but varies significantly when time backing. The orbit of NGC 4147 does not resemble that of Sgr dSph. 
\item[2).]The total orbital energy and angular momentum along the z-direction are not similar with the 
Sgr core, stream stars, or those GCs accreted from Sgr. 

\end{enumerate}

These results provide new insights into the origin of NGC 4147. We suggest that an association of 
NGC 4147 with Sgr seems to be implausible. NGC 4147 may have formed from other origins, such as a GSE accretion event or in situ. To understand the true origin of the GC population in 
the MW, more information for more stars from other dimensions, such as radial velocity or chemical abundance, is requested.

\begin{acknowledgments}
We thank the anonymous referee for the constructive and helpful comments. Jundan Nie acknowledges the support of the Beijing Natural Science Foundation (grant 
No.1232032), the National Key R\&D Program of China (grant Nos. 2021YFA1600401 and 2021YFA1600400), 
by the Chinese National Natural Science Foundation (grant No. 12373019), and the science research 
grants from the China Manned Space ProjectProject (grant Nos. CMS-CSST-2021-B03 and CMS-CSST-2021-A10). H. T. is supported by 
National Natural Science Foundation of China with grant Nos. 12103062, 12173046, and U2031143.
\end{acknowledgments}

\bibliography{ngc4147.v1}{}
\bibliographystyle{aasjournal}


\end{document}